\documentclass{article}
\usepackage{bnaic}
\pdfoutput=1
\usepackage{amsmath}
\usepackage{wasysym}
\usepackage{amsfonts}
\usepackage{verbatim}
\usepackage{graphicx}
\usepackage{amssymb}
\usepackage{enumitem}

\newcommand{\Rvs}{-.3cm}
\newtheorem{mydef}{Definition}

\title{\textbf{\huge Quantified Degrees of Group Responsibility (Extended Abstract)\footnote{The full version of this paper appears  in \cite{quantified}.}}}

\author{Vahid Yazdanpanah \and Mehdi Dastani}
\date{\textit{Utrecht University, The Netherlands}}

\begin{document}
\ttl
\thispagestyle{empty}

\section{Introduction}

This paper builds on an existing notion of group responsibility in \cite{coalitional} and proposes two ways to define the degree of group responsibility: \emph{structural} and \emph{functional} degrees of responsibility. These notions measure the potential responsibilities of (agent) groups for avoiding a state of affairs. According to these notions, a degree of responsibility for a state of affairs can be assigned to a group of agents if, and to the extent that, the group has the potential to preclude the state of affairs. 

\vspace{\Rvs}
\section {Preliminaries}

In this work, the behaviour of the multi-agent system is modelled  in a \emph{Concurrent Game Structure} (CGS) \cite{ATL} which is a tuple $M= (N,Q,Act, d, o)$, where $N=\{1, \dots ,k\}$ is a set of agents, $Q$ is a set of states, $Act$ is a set of actions, function $d:N \times Q \to \mathcal{P}(Act)$ identifies the set of available actions for each agent in $N$ at each state $q\in Q$, and $o$ is a transition function that assigns a state $q'=o(q,\alpha_1, \dots ,\alpha_k)$ to a state $q$ and an action profile $(\alpha_1, \dots ,\alpha_k)$ such that all $k$ agents in $N$ choose actions in the action profile respectively. Finally, a \emph{state of affairs} refers to a set $S \subseteq{Q}$ and  $\bar{S}$ denotes the set $Q\setminus S$. In the rest of this paper, we say $C\subseteq{N}$ is (weakly) $q$-responsible for $S$ iff it can preclude $S$ in $q$ (see \cite{coalitional} for formal details).

Let $M$ be a multi-agent system, $S$ a state of affairs in $M$, $C\subseteq{N}$ an arbitrary group, and $\hat{C}$ be a (weakly) $q$-responsible for $S$ in $M$.

\begin{mydef}[Power measures]\label{def:one}
We say that the structural power difference of $C$ and $\hat{C}$ in $q \in Q$ with respect to $S$ in $M$, denoted by $\Theta_q^{S,M}(\hat{C},C)$, is equal to cardinality of $\hat{C}\backslash C$. Moreover, we say that $C$ has a power acquisition sequence $\langle \bar{\alpha_1}, \dots ,\bar{\alpha_n} \rangle$ in $q\in Q$ for $S$ in $M$ iff for $q_i \in Q$, $o(q_i,\bar{\alpha_i})=q_{i+1}$ for  $1\leq i \leq n$ such that $q=q_1$ and $q_{n+1}=q'$ and $C$ is (weakly) $q'$-responsible for $S$ in $M$.
\end{mydef}

\vspace{\Rvs}
\section{Structural Degree of Responsibility}

In  our conception of \emph{Structural Degree of Responsibility} ($\mathcal{SDR}$), we say that any (agent) group that shares members with the responsible groups, should be assigned a degree of responsibility that reflects its proportional contribution to the responsible groups. Accordingly, the relative size of a group and its share in the responsible groups for the state of  affairs are substantial parameters in our formulation of the structural responsibility degree. We would like to emphasize that this concept of responsibility degree is supported by the fact that beneficiary parties, e.g., lobbyists in the political context, do proportionally invest their limited resources on  the groups that can play a role in some key decisions.  

\begin{mydef}[Structural degree of responsibility] \label{def:SDR}
Let $\mathbb{W}_q^{S,M}$ denote the set of all (weakly) $q$-responsible groups for state of affairs $S$ in multi-agent system $M$, and $C\subseteq{N}$ be an arbitrary group. In case $\mathbb{W}_q^{S,M}=\varnothing$, the structural degree of $q$-responsibility of any $C$ for $S$ in $M$ is undefined; otherwise, the structural degree of $q$-responsibility of $C$ for $S$ in $M$ denoted $\mathcal{SDR}_q^{S,M}(C)$, is defined as follows: 

$$\mathcal{SDR}_q^{S,M}(C)=\max\limits_{\hat{C}\in \mathbb{W}_q^{S,M}}(\{i \mid i=1-\frac{\Theta_q^{S,M}(\hat{C},C)}{\mid \hat{C} \mid}\})$$
\end{mydef}

Intuitively, $\mathcal{SDR}_q^{S,M}(C)$ measures the highest contribution of a group $C$ in a (weakly) $q$-responsible $\hat{C}$ for $S$. Hence, structural degree of responsibility is in range of $[0,1]$. 

\vspace{\Rvs}
\section{Functional Degree of Responsibility}

\emph{Functional Degree of Responsibility} ($\mathcal{FDR}$) addresses the dynamics of preclusive power of a group of agents (in the sense of \cite{power}) with respect to a given state of affairs. We deem that a reasonable differentiation could be made between the groups which do have the chance of acquiring the preclusive power and those they do not have any chance of power acquisition. This notion addresses the eventuality of a state in which a group possesses the preclusive power regarding the state of affairs. This degree is formulated based on the notion of \emph{power acquisition sequence} (Definition \ref{def:one}) by tracing the number of necessary state transitions from a source state, in order to reach a state in which the group in question is responsible for the state of affairs. 

\begin{mydef}[Functional degree of responsibility]\label{def:fdr}
Let $\mathbb{P}_q^{S,M}(C)$ denote the set of all power acquisition sequences of $C\subseteq{N}$ in $q$ for $S$ in $M$. Let also $\ell$ = $\min\limits_{k \in \mathbb{P}_q^{S,M}(C)}  (\{ i \mid i=length(k)  \})$ be the length of a shortest power acquisition sequence. The functional degree of $q$-responsibility of $C$ for $S$ in $M$, denoted by $\mathcal{FDR}_q^{S,M}(C)$, is defined as follows:
\[ \mathcal{FDR}_q^{S,M}(C) = \left\{
  \begin{array}{l l}
    0 & \quad \text{if }\ \mathbb{P}_q^{S,M}(C)=\varnothing\\
    \frac{1}{(\ell+1)} & \quad \text{otherwise}
  \end{array} \right.\]
\end{mydef}

The notion of $\mathcal{FDR}_q^{S,M}(C)$ is formulated based on the minimum length of power acquisition sequences, which taken to be  $0$ if $C$ is a (weakly) $q$-responsible for $S$. Hence, the functional degree of $q$-responsibility of such a $C$ for $S$ is equal to $1$. If there exists no power acquisition sequence for $C$, then the minimum length of a power acquisition sequence is taken to be $\infty$ and the functional degree of  $q$-responsibility of $C$ for $S$ becomes $0$. In other cases $\mathcal{FDR}_q^{S,M}(C)$ is strictly between $0$ and $1$. 

\vspace{\Rvs}
\section{Conclusion}\label{Sec:Conc}

The proposed notions can be used as a tool for analyzing the potential responsibility of agent groups towards a state of affairs. In our approach, the \emph{structural degree of responsibility} captures the responsibility of an agent group based on the accumulated preclusive power of the included agents while the \emph{functional degree of responsibility} captures the responsibility of a group of agents due to the potentiality of reaching a state in which it has the preclusive power. In the full version of the paper, we specify pertinent properties of the notions and consider additional semantics.

\vspace{\Rvs}
\bibliographystyle{plain}
\bibliography{01_Main}
\end{document}